\documentclass[twocolumn,prx,superscriptaddress,preprintnumbers,amsmath,amssymb]{revtex4-2}
\usepackage{graphicx}
\usepackage{color}

\begin{document}

\title{Symmetry breaking in two dimensions on ultra-fast time scales}

\author{Alireza Valizadeh}
\affiliation{Max-Planck-Institute for Dynamics and Self-Organization, 37077 G\"ottingen, Germany}
\affiliation{Institute for the Dynamics of Complex Systems, University of G\"ottingen, 37077 G\"ottingen, Germany}

\author{Patrick Dillmann}
\affiliation{Department of Physics, University of Konstanz, D-78467 Konstanz, Germany}
\author{Peter Keim}

\email{peter.keim@uni-duesseldorf.de}
\affiliation{Institute for Experimental Physics of Condensed Matter, Heinrich-Heine-Universit\"at D\"usseldorf, 40225 D\"usseldorf, Germany}
\affiliation{Institute for the Dynamics of Complex Systems, University of G\"ottingen, 37077 G\"ottingen, Germany}
\affiliation{Max-Planck-Institute for Dynamics and Self-Organization, 37077 G\"ottingen, Germany}

\date{\today}

\begin{abstract}

Melting of two-dimensional mono-crystals is described within the  celebrated Kosterlitz-Thouless-Halperin-Nelson-Young scenario (KTHNY-Theory) by the dissociation of topological defects \cite{Kosterlitz1972,Kosterlitz1973,Nelson1977,Halperin1978,Young1979}. It describes the shielding of elasticity due to thermally activated topological defects until shear elasticity disappears. As a well defined continuous phase transition, freezing and melting should be reversible and independent of history. However, this is not the case: cooling an isotropic 2D fluid with a finite but nonzero rate does not end in mono-crystals. The symmetry can not be broken globally but only locally in the thermodynamic limit due to the critical slowing down of order parameter fluctuations. This results in finite sized domains with the same order parameter. For linear cooling rates, the domain size is described by the Kibble-Zurek mechanism, originally developed for the defect formation of the primordial Higgs-field shortly after the
Big-Bang. In the present manuscript, we investigate the limit of the deepest descent quench on a colloidal monolayer and resolve the time dependence of structure formation for (local) symmetry breaking. Quenching to various target temperatures below the melting point (deep in the crystalline phase and just close to the transition), we find universal behaviour if the timescale is re-scaled properly. 
\end{abstract}

\maketitle

\section{Introduction}

The glass transition temperature depends on the cooling rate and shows memory effects, thus it does not mediate between well defined thermodynamic states in the classic sense \cite{Ritland1954,Vollmayr1996,Angell2000,Debenedetti2001}. This is different for phase transitions where melting and freezing is assumed to be reversible. However, this is partly true for discontinuous (first order) phase transitions, where a nucleation barrier has to be overcome. For careful cooling / heating and very clean ensembles (free of nucleation seed leading to heterogeneous nucleations) the systems can be supercooled or overheated, indicating some hysteresis in transition temperatures. 
The maximal width of the hysteretic region ranges up to the spinodal, where the slope of e.g. Van-der Waals curve becomes negative and phase separation must set in \cite{Binder1987}.\\

For continuous (second-order) phase transitions, nucleation barriers do not exist, ruling out any hysteresis of the transitions. Since the free energy density between the high and low temperature phase disappears at the transition temperature, phase separation does not exist and there is a priory no reason to expect freezing and melting not to be completely reversible. However, critical fluctuations of the order parameter dictate the behaviour of the ensemble and timescales become important. A famous example is the Ising model in 2D which is beside the 2D XY-model and the 2D particle system one of the rare examples with an analytic solution of transitions temperatures $T_c$ \cite{Onsager1944,Kosterlitz1972,Kosterlitz1973,Halperin1978,Nelson1979,Young1979}. Above $T_c$ magnetic moments randomly point upwards and downwards, the ensemble is homogeneous and isotropic. Below $T_c$, the magnetic moments organize to be parallel and a macroscopic magnetization builds up. This magnetization can be taken as an order parameter and the symmetry of the ensemble is broken. If the magnetization points upwards or downwards is a matter of change - part of the story why this phenomenon is called \emph{spontaneous} symmetry breaking. Typically it is assumed that by crossing $T_c$ from the high temperature to the low temperature phase the symmetry switches globally. However, taking timescales into account this can not be the truth in the thermodynamic limit: in infinitely large ensembles, only regions which are connected by causality can have the same order parameter after symmetry breaking. Regions which are separated a distance larger than the speed of light times the time after crossing $T_c$ (defining an event horizon) can not necessarily gain the same order parameter. Thus, the symmetry can only be broken locally and the order parameter can be uniform only within the event horizon. Melting and freezing are not reciprocal in the thermodynamic limit, even for continuous phase transitions.\\

This idea was first discussed by Tom Kibble \cite{Kibble1976,Kibble1980,Kibble2007} for the symmetry breaking of the first two component scalar field (nowadays often named Inflaton) shortly after Big-Bang. Topological defects like grain boundaries, strings, and monopoles should be incorporated as leftovers of the high symmetry field in the symmetry-broken stage. Zel'dovich e.g. calculated optical properties of grain boundaries within the Higgs-field for detecting their traces within the electromagnetic background radiation in the universe \cite{Zeldovich1974}. Up to now, such defects have never been observed which is one of the reasons (besides the flatness of space-time of the universe and the overall very isotropic cosmic microwave background) postulating inflationary Big Bang models: during the exponentially fast growth of the very early universe, all defects have been pushed beyond the event horizon.\\

\begin{figure}
  \includegraphics[width=1\linewidth]{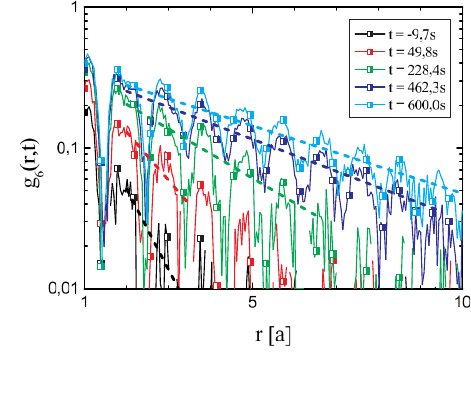}
 \caption{\label{fig:g6r} Bond-order correlation function before the quench (black) and after a quench from $\Gamma=16$ to a final coupling strength of $\Gamma_E=74$ for various waiting times. Dotted lines are exponential fits to the curves, which appear as straight lines in the lin-log-plot.}
\end{figure}

Wojciech Zurek \cite{Zurek1985,Zurek1993,Zurek1996} applied the same idea to another two-component but complex field, namely the macroscopic wave function of superfluid Helium. In this three-dimensional (3D) system with a two-component (2N) order parameter, the most natural topological defects are strings, given by the vortices of the wave function with a normal-fluid core. In condensed matter, the signal velocity is given by the speed of sound (or second sound in the case of superfluid Helium). However, a more detailed picture of defect formation is as follows: any continuous phase transition is dominated by critical fluctuations. Approaching $T_c$ from above, ordered domains get larger and larger and order parameter correlation functions diverge algebraically (free of any typical length-scale). At $T_c$, the correlation length is infinite, the structure is self-similar on all scales showing a fractal pattern. Both, ordered and disordered regions cover $50\%$ of the volume. Since the energy difference between both phases is zero at $T_c$ (and very small in the vicinity), the pattern is not static but fluctuates. The larger the domains, the slower they appear and disappear. Not only length scales diverge but also time scales: the temporal order parameter correlation functions diverge algebraically, too. This behaviour in the vicinity of $T_c$ is named critical slowing down.\\

For any nonzero (most easily linear) cooling rate, one can compare the critical slowing down with the time to reach the transition. Far away from the transition, correlation times are short and order-parameter fluctuations can follow the cooling. The system is quasi-adiabatic. If the correlation time gets larger than the time to reach the transition, fluctuations can not follow further and a fingerprint of the longest length scale is taken. This well-defined fall-out-time defines the largest size of the symmetry-breaking domains. Many systems have been investigated to prove the Kibble-Zurek mechanism, e.g. in liquid crystals~\cite{Chuang1991}, superfluid $^3$He~\cite{Bauerle1996}, superconducting systems~\cite{Carmi2000}, convective, intrinsically out of equilibrium systems~\cite{Miranda2013}, multiferroics~\cite{Chae2012}, quantum systems~\cite{Xu2014}, ion crystals~\cite{Ulm2013,Pyka2013}, and Bose-Einstein condensates~\cite{Lamporesi2013} (the latter two systems contain the effect of inhomogeneities due to e.g. temperature gradients). A detailed review concerning the significance and limitations of these experiments can be found in~\cite{Kibble2007,Campo2014}. Recent work includes quantum system also in two dimensions \cite{Keesling2019,Ebadi2021,Schmitt2022} and universal behaviour as well as limitations of the Kibble-Zurek scaling are discussed in \cite{Campo2018,Zeng2023}. In classical two dimensions ensembles and for linear cooling rates, the Kibble-Zurek scaling was proven to be valid also for the Kostelitz-Thouless universality \cite{Deutschlaender2015}.\\
\begin{figure}
  \includegraphics[width=0.9\linewidth]{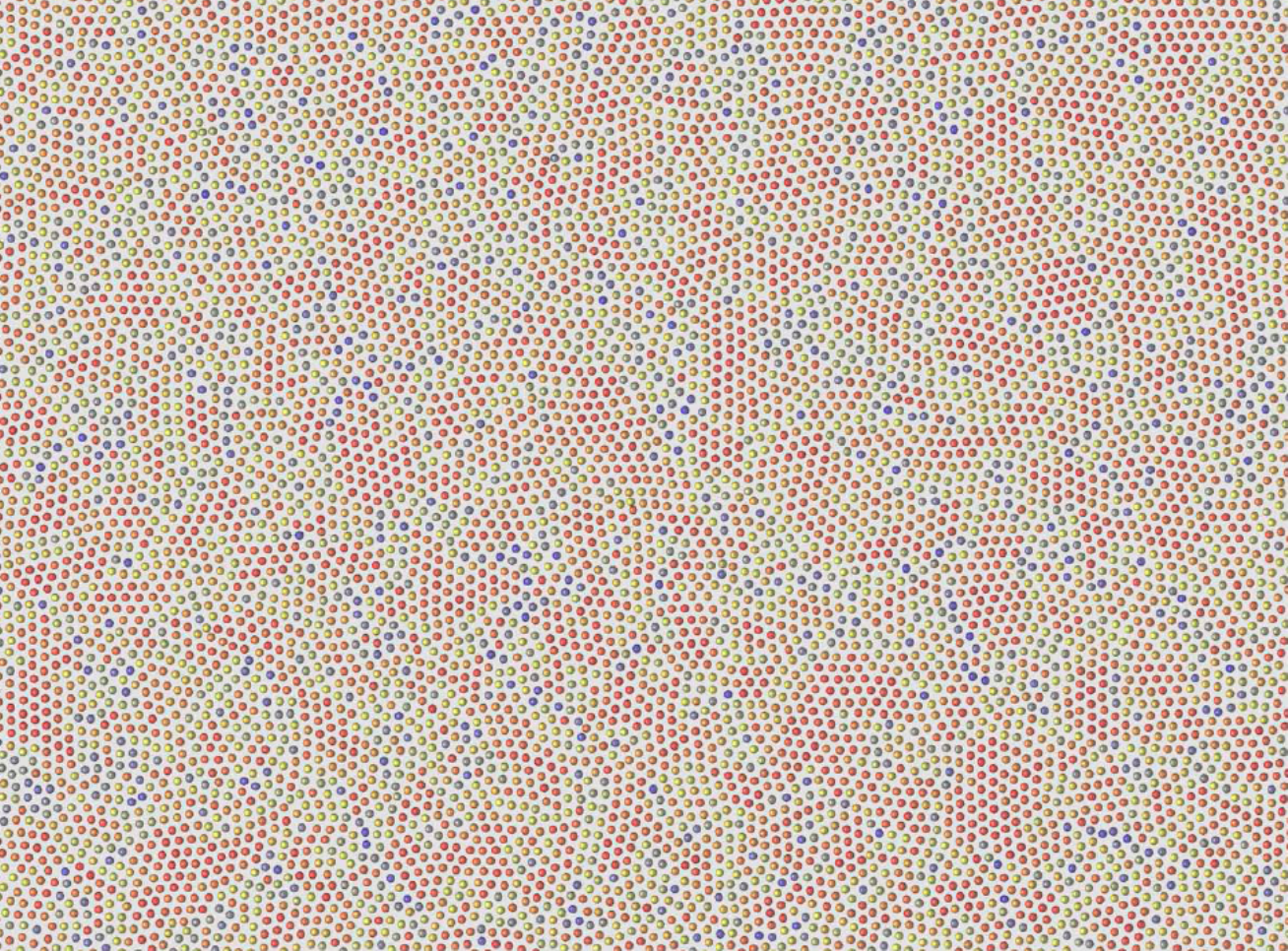}
 \caption{\label{fig:pos_color_code} Snapshot of the mono-layer $120\;\mbox{s}$ after a quench from $\Gamma=16$ to a final coupling strength of $\Gamma_E=166$. The color code (from blue $m=0$ over yellow to red $m=1$) is given by the magnitude $m$ of the local bond order parameter. Blue particles are locally disordered (low order = high symmetry phase) and red particles have large sixfold symmetry (high order = broken symmetry).} 
\end{figure}

How does the scenario alter, if the system is quenched on ultra-fast time scales? In the given manuscript we investigate a two-dimensional (2D) colloidal mono-layer which has a two-component order parameter (2N) given by the director field of nearest neighbors. This property is shared by the Higgs-field and superfluid Helium, besides the fact, that the latter is a complex quantity. Since it is two-dimensional and consists of micrometer-sized particles, defects and domains can be monitored in situ by video microscopy, unlike in 3D ensembles, where typically only the surface of the 3D bulk can be monitored.\\

\begin{figure}
\includegraphics[width=1.\linewidth]{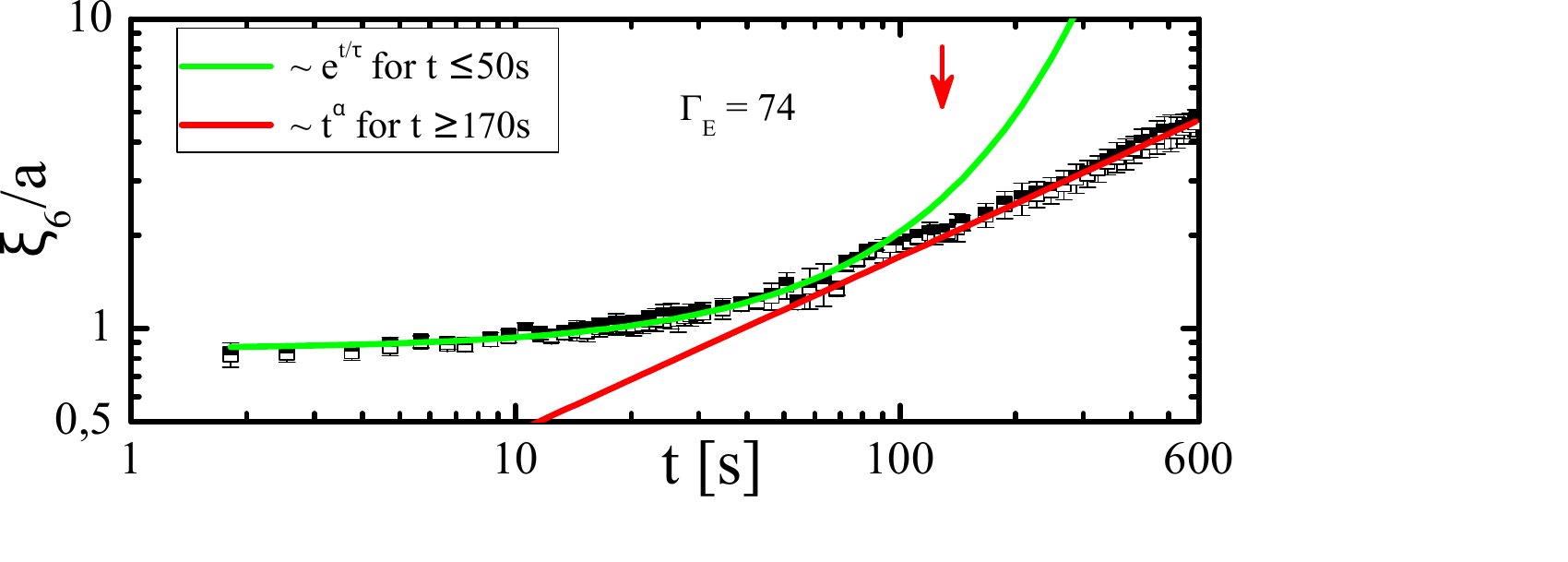}
\includegraphics[width=1.\linewidth]{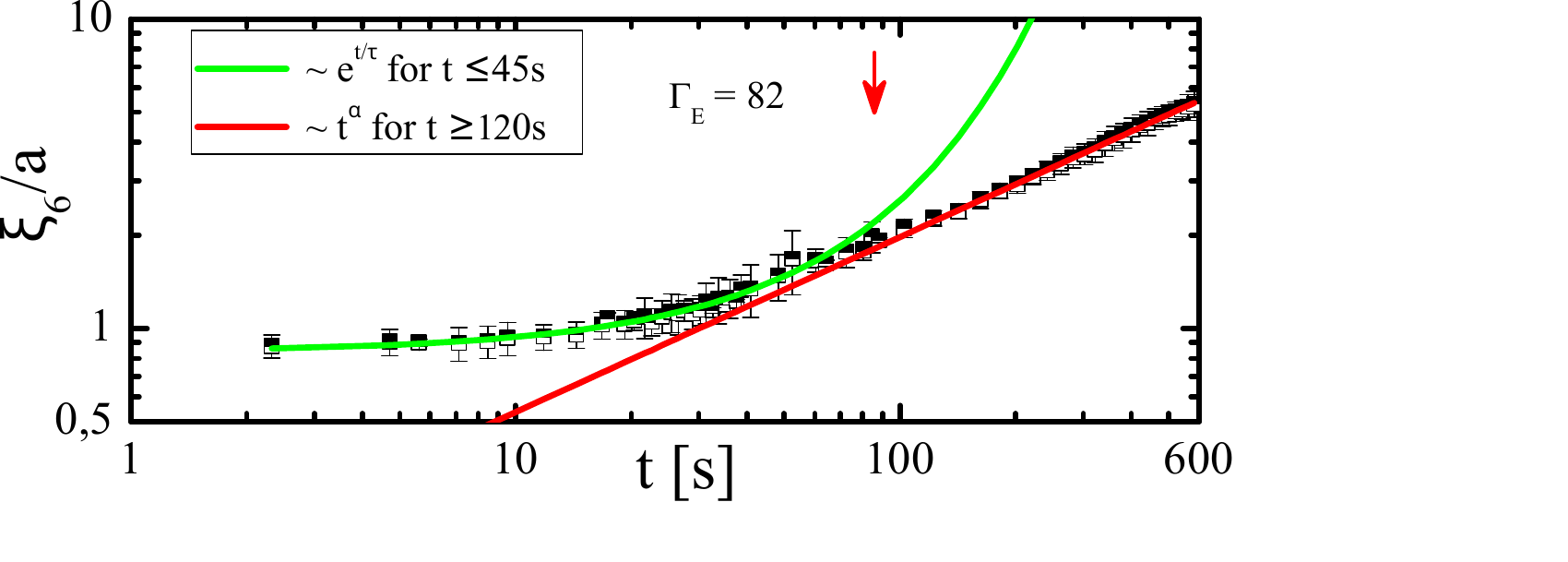}
\includegraphics[width=1.\linewidth]{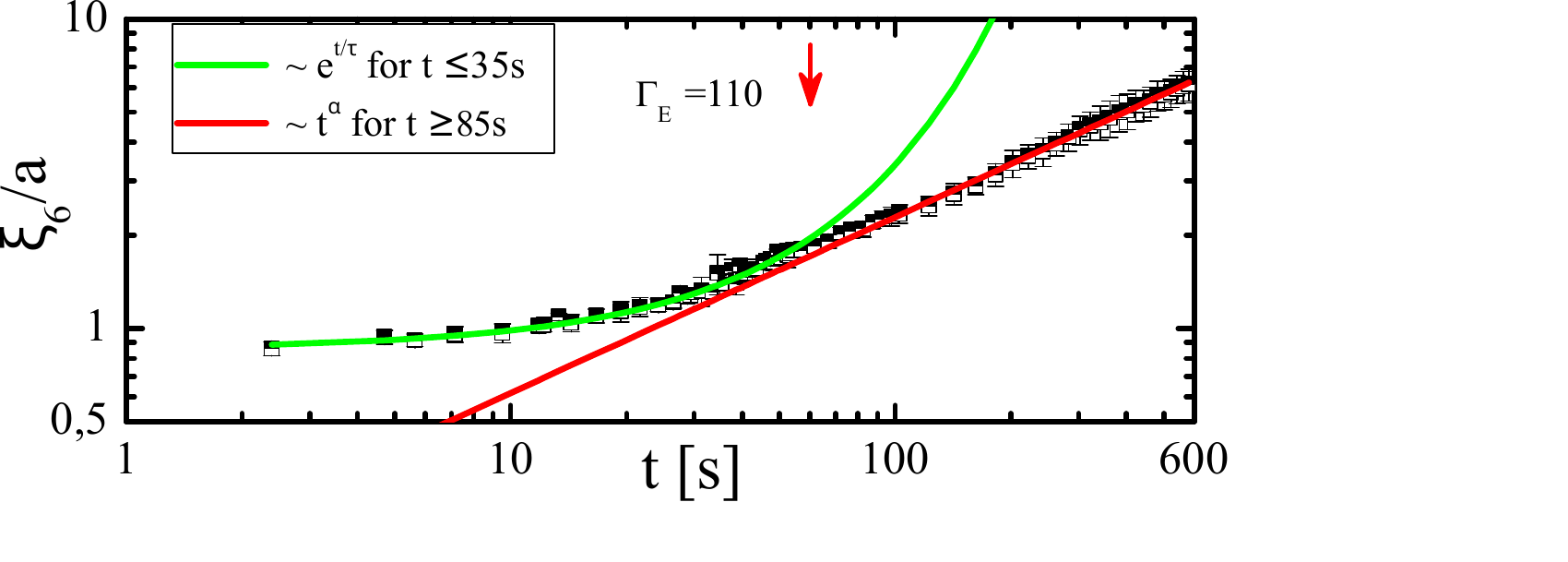}
\includegraphics[width=1.\linewidth]{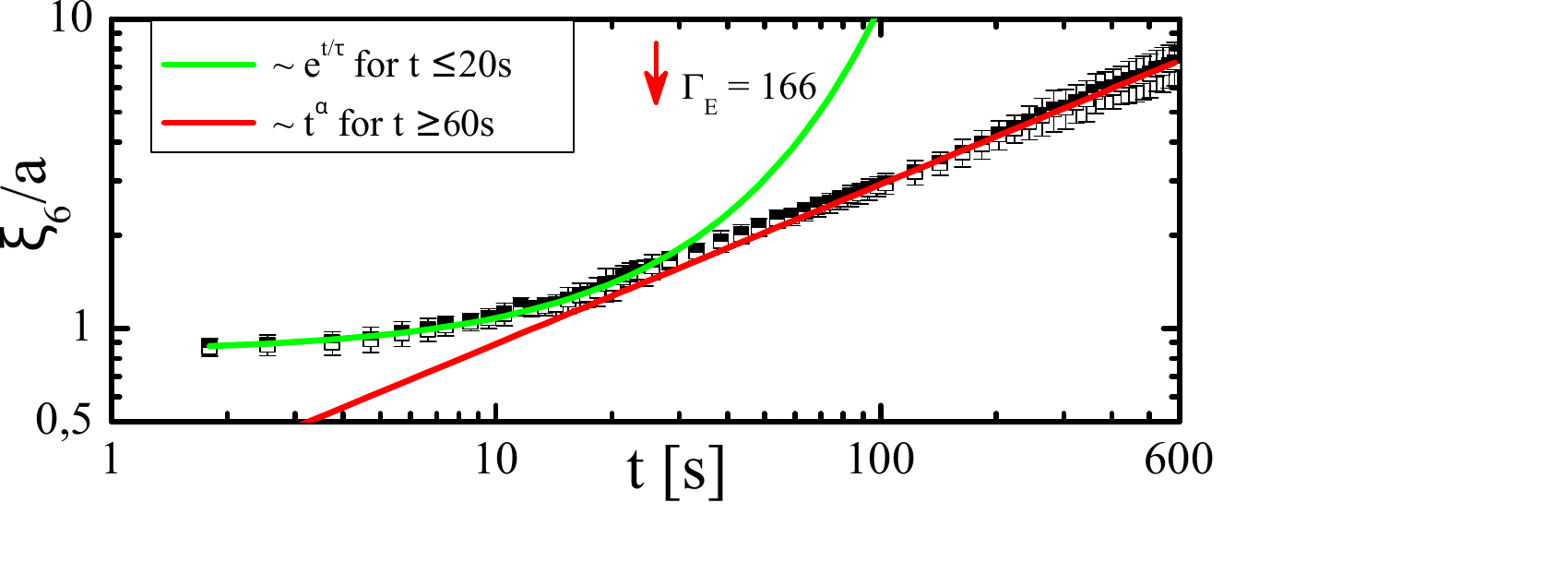}
\caption{\label{fig:xi_6(t)} Orientational correlation length as function of time in a log-log-plot for different quench depth. Shortly after the quench, the growing behavior is exponential and switches to algebraic later. The quench depth increases from top to bottom. Note the decreasing cross-over time (red arrows as a guide for the eye) for deeper quenches. The crossover from exponential to algebraic takes place when the number of domains has its maximum and clusters start to tough. This value is at about $~37\% $ crystallinity which marks the time range when fluctuations start to become suppressed.}
\end{figure}

The experiment is in detail described in \cite{ebert2009a} and it was successful in validating 2D melting theory \cite{Keim2007,Gasser2010,Kosterlitz2017} and Mermin-Wagner-Hohenberg fluctuations \cite{Illing2017}. Here, only a brief description is given: a droplet of a suspension composed of super-paramagnetic polystyrene spheres \footnote{Dynabeads, 4.5 $\mu$m, http://www.dynal.no} are dispersed in a water droplet, which is suspended by surface tension in a top sealed cylindrical hole ($\varnothing = 6\;mm$) of a glass plate. The particles are $4.5\;\mu m$ in diameter and have a mass density of $1.5\;g/cm^3$ leading to sedimentation. The large density is due to the fact that the polystyrene beads are doped with iron oxide nano-particles which further causes the super-paramagnetic behavior. After sedimentation, particles are arranged in a mono-layer at the water-air interface of the droplet. The interface is kept planar (less than $250\,\mathrm{nm}$ height difference from the middle to the border) by active regulation based on several control loops using digital image analysis. It is kept horizontal with changes in inclinations less than $1\,\mathrm{mRad}$ by an inclinations sensor driving a tripod on which the whole setup is mounted. This way, the ensemble forms an ideal 2D system, without any pinning of particles to the substrate. The particles themselves are small enough to perform 2D Brownian motion but large enough to be monitored with video microscopy. The field of view of the video camera is $1160 \times 865\;\mu m^2$ in size and contains about $9000$ colloids, while the whole mono-layer contains $\sim 300\,000$ particles. An external magnetic field $H$ perpendicular to the water-air interface induces a magnetic moment in each bead (parallel to the applied field) leading to repulsive dipole-dipole interaction between all particles. We use the dimensionless control parameter $\Gamma$ to characterize this interaction strength. $\Gamma$ is given by the ratio of dipolar magnetic energy versus thermal energy 
\begin{equation} \Gamma=\frac{\mu_0}{4\pi}\frac{(\chi H)^2(\pi \rho)^{3/2}}{k_BT} \propto T_{sys}^{-1} 
\end{equation} 
and thus can be regarded as a dimensionless inverse system temperature or a dimensionless in-plane pressure. The state of the system in thermal equilibrium - liquid, hexatic, or solid - is solely defined by the strength of the magnetic field $H$ since the laboratory temperature $T$, the 2D particle density $\rho$ and the magnetic susceptibility per bead $\chi$ are kept constant experimentally.\\

In these units the transition (crystalline - hexatic) is at $\Gamma_m = 70\pm 0.5$ and the transition from hexatic to isotropic at about $\Gamma_i = 68\pm 0.5$ \footnote{Earlier work \cite{Keim2007} reports lower values due to a lower magnetic susceptibility of the colloids, measured by SQUID. Meanwhile, it is more precise to determine the susceptibility in comparison of the pair correlation function in the fluid phase with computer simulations, leading to the given transition temperatures.}. Since the system temperature is given by an outer field, enormous cooling rates are accessible compared to atomic systems. Based on a well-equilibrated liquid system 
at $\Gamma \approx 16$, deep in the fluid phase, we initiate a temperature jump with cooling rates up to $ d\Gamma /dt \approx 10^4~s^{-1}$ into the crystalline region of the phase diagram $\Gamma_m \geq 70$. This temperature quench triggers the solidification within the whole mono-layer. The time scale of cooling is $10^5$ faster compared to the fastest intrinsic scales, given e.g. by the Brownian time $\tau_B = 50\;\mathrm{sec}$, which is the time a particle needs to diffuse the distance of it's own diameter. In atomic systems this time scale is much faster (Brownian time in water is $ \sim 10^{-11}\;\mathrm{sec}$) thus comparable quench rates are rarely accessible. \\

Another enormous advantage of the two-dimensional ensemble is, that there is no heat flux through the bulk and especially through the surface of the material as it is usually the case in 3D condensed matter systems. Especially at extreme cooling rates, this would easily lead to temperature or pressure gradients causing an inhomogeneous background during symmetry breaking. This is ruled out in our experiment. To increase statistics, each temperature quench is repeated at least ten times to the same designated value of the control parameter $\Gamma_{end}$ with sufficient equilibration times in between.\\

\section{Results}
\subsection{Orientational correlations}

Ordering in 2D is best measured with the six-folded bond order correlation function $g_6(r)$, since the closed-packed crystal structure in 2D is always hexagonal, at least for isotropic interaction:
\begin{equation}
\label{g6} g_6(r) = \langle |\psi(\vec{r_k})\psi^\ast(\vec{r_j})|
\rangle_{kj} = \langle |\psi(\vec{r})\psi^\ast(\vec{0})| \rangle
\quad ,
\end{equation}
$g_6(r)$ is based on the local director field given by the nearest neighbors. For a particle numbered with $l$ it reads $\psi_l = 1/ N_j \sum_{k=1}^{N_j} e^{i 6\theta_{kl}}$. Here, $N_j$ counts the nearest neighbors which define the bond directions $\theta_{kl}$ with respect to a fixed but arbitrary axis. Fig~\ref{fig:g6r} shows the bond order correlation function $g_6(r)$ for one time before and various times after the quench in a lin-log plot. For all times it decays exponentially and no signature of algebraic decay is detectable. Thus the orientational order is always short range and no quasi-long range order appears, ruling out signatures of a hexatic phase, being prominent in 2D melting \cite{Keim2007}. To have a closer look we plot the positions of particles and measure the local order: Fig \ref{fig:pos_color_code} 
shows the mono-layer two minutes after the quench from $\Gamma_{start}\approx16$ to $\Gamma_{end}\approx166$. The color code of particle $l$ at position $\vec{r}_l$ is given by the magnitude $ m_l = m(\vec{r_l}) = \psi_l^\ast\psi_l$ of the local complex (six-folded) bond order field $\psi_l$. What is observed are patches of sixfold order but with different orientations of the director separated by disordered regions. Drawing a close analogy of poly-crystallinity, we introduce the term 'poly-hexallinty' to describe such systems. In equilibrium, the hexatic phase is defined by an algebraic decay of the bond order correlation function but only in the asymptotic limit of large distances. Out of equilibrium, the Kibble-Zurek mechanism causes the symmetry to be broken only locally thus the poly-hexatic phase will always lead to a decay of the bond order correlation function faster than algebraic. However, locally symmetry broken domains are better described by their orientational order compared to translational order. Since the latter is in principle not broken in two dimensions due to the Mermin-Wagner-Hohenberg theorem \cite{Mermin1966,Hohenberg1967,Mermin1968,Illing2017} (even in a mono-crystal without having poly-crystallinity) we proceed with bond order correlations.\\ 

\begin{figure}
 \includegraphics[width=1.\linewidth]{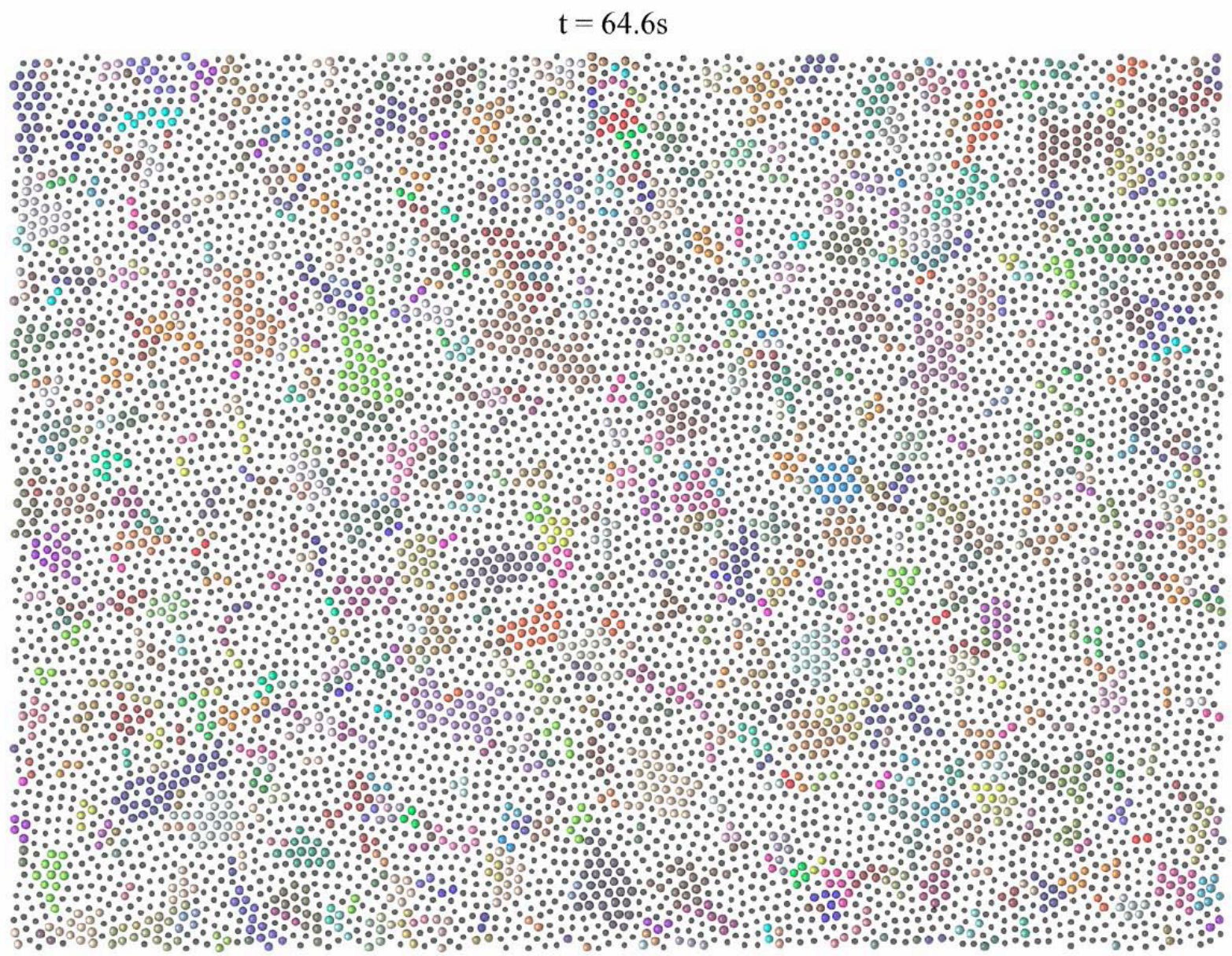}
 \caption{\label{fig:sym_broken_domains} Snapshot of the monolayer with symmetry broken domains (SBD), $64\;\mbox{s}$ after a quench from $\Gamma=13$ to a final coupling strength of $\Gamma_E=110$. Small dots are fluid-like particles, whereas big ones are crystal-like. Different colors indicate individual grains which are labeled in time.}
\end{figure}

From the exponential decay $g_6(r,t) \sim \exp(-r/\xi_6(t)) $ of the bond order correlation length (Fig.~\ref{fig:g6r}) one can extract the orientational correlation length $\xi_6(t)$ as function of time for different quench rates. As shown in Figure~\ref{fig:xi_6(t)} the correlation length always grows monotonically after the quench. This is true for low quenches ($\Gamma_F = 74$), intermediate ($\Gamma_F = 80~\mathrm{ and }~110$), and deep quenches ($\Gamma_F = 166$). 
In the early stage, we observe a non-algebraic growth of the domain size (green curves) followed by an algebraic one (red curves) for all quench depth. Excluding the crossover region (as given in the label) the non-algebraic time window is best fitted with an exponential increase of bond order correlation length. In principle $\xi_6(t)$ is a measure for the average size of symmetry broken domains but comparing Fig.~\ref{fig:pos_color_code} where the domains extend about half a dozens particles in diameter, the correlations length in Fig.~\ref{fig:xi_6(t)} (lowest plot) after $120~\mathrm{sec}$, where the ensemble is already in the algebraic time window, is about three particle distances. The residuum of the high symmetry phase affects the averaged bond order significantly. Thus, we introduce another criterion to measure the size of locally symmetries broken domains that furthermore allows us to follow and label individual domains in time: simply taking a threshold for the magnitude of the local bond order orientation is not sufficient and an elaborate discussion how to define crystallinity in 2D on a local scale can be found in \cite{Dillmann2013}. \\ 

\begin{figure*}
\centering
\begin{tabular}{cc}
\includegraphics[width=0.47\linewidth]{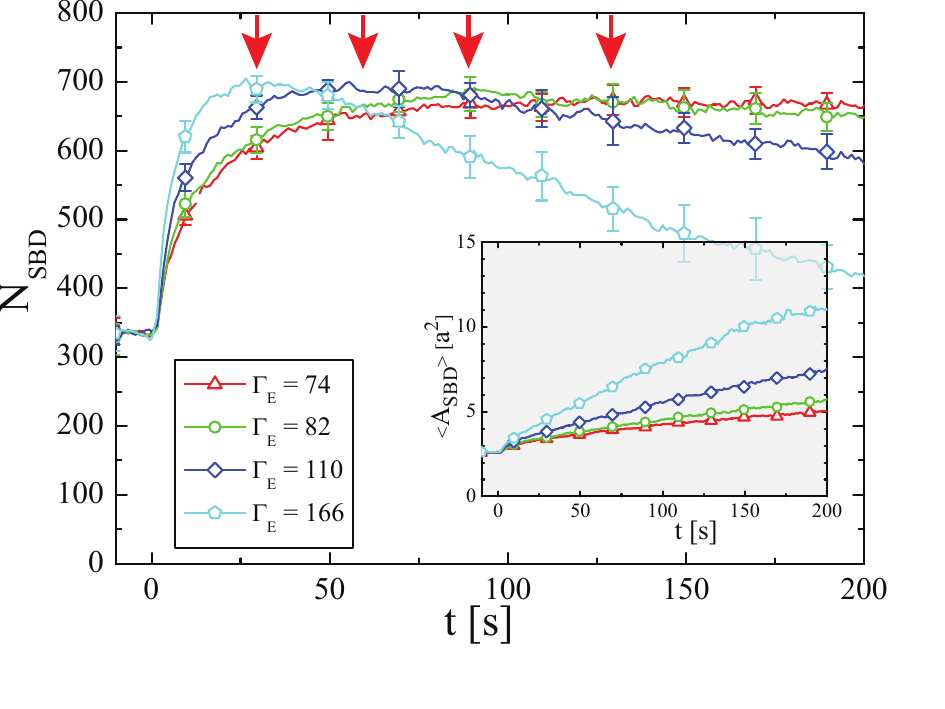} & \includegraphics[width=0.47\linewidth]{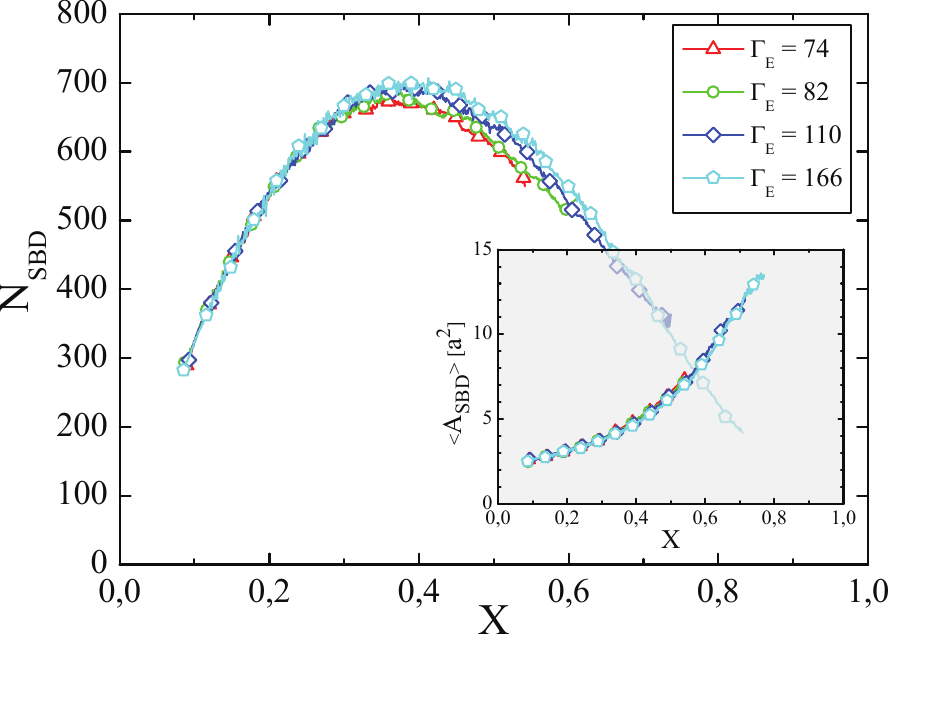} \\
\end{tabular}
\caption{\label{fig:NuA(ToX)} The left image shows the number of symmetry-broken domains (SBD) and the average size of the domains (inset) as a function of time for different quench depths. The error bars are averages about $10$ independent quenches with about $9000$ particles in the field of view. Whereas the average size of the nuclei grows monotonically as expected, the number of symmetry-broken domains first increases to a maximum but then decreases in favor of fewer but larger domains. The red arrows are a guide to the eye to identify the maxima. The right image shows the same data but plotted as the fraction of symmetry-broken area which is implicitly a function of time. The curves almost superimpose as a function of crystallinity and are independent of the quench depth.}
\end{figure*}

\subsection{Determining symmetry broken domains}

We define a particle to be part of a symmetry-broken domain if the following three conditions are fulfilled for the particle itself and at least one nearest neighbor:
\begin{itemize}
\item The magnitude of the local bond order field $m_{6_k}$ must exceed $0.6$ for both neighboring particles.
\item The bond length deviation $\Delta |l_{kl}|$ of neighboring particles $k$ and $l$ is less than $10\%$ of the average particle distance $l_a$.
\item The variation in bond orientation $\Delta\Theta = |\psi_k-\psi_l |$ of neighboring particles $k$ and $l$ must be less than $2.3^\circ$ in real space (less than $14^\circ$ in six-folded space).
\end{itemize}
Simply connected domains of particles that fulfill all three criteria are merged into a local symmetry-broken domain. \\
Figure~\ref{fig:sym_broken_domains} shows a snapshot of the ensemble with local domains marked in different colors, while particles still in a high-symmetry configuration are plotted as small gray dots. From this data, one can analyze the number and size of domains as a function of time.\\

\subsection{Number and size of domains}

Figure~\ref{fig:NuA(ToX)} [left] shows the average number of local domains as a function of time up to $200\;\mbox{sec}$ after the quench and the inset shows the mean size. As expected, the mean size of the nuclei grows monotonically as a function of time for all quench depths. The average number of domains (Fig.~\ref{fig:NuA(ToX)} [left]) first increases but finally decreases in time. The maximum shifts to shorter times as a function of quench depth - deeper supercooling drives the system faster to the solid state. Note, that the growing of domains starts immediately after the quench such that no lag time known from classical nucleation theory (CNT) is detectable. Figure~\ref{fig:NuA(ToX)} [right] shows the same data but plotted as a function of crystallinity $X$ instead of time. Here, we define crystallinity $X$ as the fraction of particles belonging to a symmetry-broken domain with respect to all particles. Interestingly all curves almost superimpose and show universal behaviour as a function of crystallinity. Surprisingly, the position and height of the average number of symmetry-broken domains (in the context of nucleation this is called the mosaicity) is independent of the quench depth. It appears when roughly $37\%\sim 1/e$ of particles belong to symmetry-broken domains. Comparing the maximum on the real-time axis (red arrows in Fig.~\ref{fig:NuA(ToX)} [left]) and the cross-over time in Fig.~\ref{fig:g6r} for different quench depth we propose the following scenario: After a quench, local symmetry-broken domains start to grow exponentially until about $37\%$ of the space is covered. In this time window, the ensemble is dominated by critical-like fluctuations and most of the domains disappear again. In our 2D system, the fraction of $37\%$ crystallinity marks a threshold where domains with different orientations start to touch. From this on, critical-like fluctuations are suppressed and the following dynamics is dominated by the conversion of the yet untransformed regions of the high symmetry phase in the region between symmetry broken domains. This period is marked by an algebraic increase in the bond order correlation length. For finite cooling rates, G. Biroli et al. have argued \cite{Biroli2010} that defect annihilation will alter the classical Kibble-Zurek mechanism leading to two different time regimes after the fall-out time, separating critical and classical coarse-graining. We argue that at very early times and quenches from a rather hot true vacuum (in terms of high energy physics), two also different time regimes are detectable, but less affected by annihilation of defects but critical-like fluctuations before and after domains get in touch.

\section{Conclusion}

The paradigmatic melting theory in 2D, the Kosterlitz-Thouless-Halperin-Nelson-Young-theory starts with a highly idealized mono-crystal, free of dislocations, disclinations, vacancies, interstitials, or grain boundaries. Recursion relations for elasticity, shielded by defects can be solved by renormalization group theory until shear elasticity disappears under heating, entering a fluid phase. However, this mono-crystal can never be obtained by cooling, at least in the thermodynamic limit. The framework is given by Kibble and Zurek, arguing with causality and critical slowing down, respectively. Out of equilibrium and for any nonzero cooling rate, the ensemble will become poly-crystalline with a domain size given by the cooling rate. In the thermodynamic limit, freezing and melting is not reversible, taking time scales into account.\\

Here, we use a mono-layer of super-paramagnetic colloids which can be cooled free of gradients on unrivaled fast time scales compared to any intrinsic dynamics of the ensemble. We find universal behaviour of the domain size and number of domains, independent of how deep we quench into the crystalline phase. This shows up most easily when the timescale is normalized by the crystallinity, the area fraction of symmetry broken domains. This is in stark contrast to ensembles with underlying first-order transitions, where the mosaicity depends on the degree of supercooling after the quench. For continuous transitions, we propose that mosaicity depends on the temperature \textit{before} the quench and one picks out the onset of criticality surprisingly far away from the transition.\\

Taking the time-dependent bond order correlation length we can identify two regimes: first an exponential growth, followed by an algebraic one. Note, that the exponential growing is not yet observed so far and we keep it open if is due to the Kosterlitz-Thouless universality of 2D ensembles. For the Lambda-transition of $He^4$ Zurek proposed an algebraic decrease of the inverse defect density after the quench \cite{Zurek1993}. For the XY-model a power-law increase of the correlation length with a small logarithmic correction is predicted \cite{Jelic2011} and in dusty plasma, an algebraic increase was found experimentally \cite{Hartmann2010}. The end of the crossover region between exponential to algebraic is at about $37\% \sim 1/e$ transformed area (or crystallinity). This coincides with the time when the number of domains starts to decrease. Our microscopic picture is as follows: 
In the exponential regime, symmetry-broken domains appear but are still strongly affected by critical-like fluctuations: most of them disappear again. At about $37\%$ crystallinity, symmetry-broken domains start to touch and critical-like fluctuations are damped. In the algebraic regime, fewer domains dissolve, larger ones more dominantly grow and the total number of domains decreases.

\begin{acknowledgments}
P.K. acknowledges financial support from German Research Foundation (DFG), project number 453041792 (Heisenberg-funding), project number 408261333, and project number237282255.
\end{acknowledgments}

\bibliography{literatur.bib}

\begin{thebibliography}{49}%
\makeatletter
\providecommand \@ifxundefined [1]{%
 \@ifx{#1\undefined}
}%
\providecommand \@ifnum [1]{%
 \ifnum #1\expandafter \@firstoftwo
 \else \expandafter \@secondoftwo
 \fi
}%
\providecommand \@ifx [1]{%
 \ifx #1\expandafter \@firstoftwo
 \else \expandafter \@secondoftwo
 \fi
}%
\providecommand \natexlab [1]{#1}%
\providecommand \enquote  [1]{``#1''}%
\providecommand \bibnamefont  [1]{#1}%
\providecommand \bibfnamefont [1]{#1}%
\providecommand \citenamefont [1]{#1}%
\providecommand \href@noop [0]{\@secondoftwo}%
\providecommand \href [0]{\begingroup \@sanitize@url \@href}%
\providecommand \@href[1]{\@@startlink{#1}\@@href}%
\providecommand \@@href[1]{\endgroup#1\@@endlink}%
\providecommand \@sanitize@url [0]{\catcode `\\12\catcode `\$12\catcode
  `\&12\catcode `\#12\catcode `\^12\catcode `\_12\catcode `\%12\relax}%
\providecommand \@@startlink[1]{}%
\providecommand \@@endlink[0]{}%
\providecommand \url  [0]{\begingroup\@sanitize@url \@url }%
\providecommand \@url [1]{\endgroup\@href {#1}{\urlprefix }}%
\providecommand \urlprefix  [0]{URL }%
\providecommand \Eprint [0]{\href }%
\providecommand \doibase [0]{https://doi.org/}%
\providecommand \selectlanguage [0]{\@gobble}%
\providecommand \bibinfo  [0]{\@secondoftwo}%
\providecommand \bibfield  [0]{\@secondoftwo}%
\providecommand \translation [1]{[#1]}%
\providecommand \BibitemOpen [0]{}%
\providecommand \bibitemStop [0]{}%
\providecommand \bibitemNoStop [0]{.\EOS\space}%
\providecommand \EOS [0]{\spacefactor3000\relax}%
\providecommand \BibitemShut  [1]{\csname bibitem#1\endcsname}%
\let\auto@bib@innerbib\@empty
\bibitem [{\citenamefont {Kosterlitz}\ and\ \citenamefont
  {Thouless}(1972)}]{Kosterlitz1972}%
  \BibitemOpen
  \bibfield  {author} {\bibinfo {author} {\bibfnamefont {J.~M.}\ \bibnamefont
  {Kosterlitz}}\ and\ \bibinfo {author} {\bibfnamefont {D.~J.}\ \bibnamefont
  {Thouless}},\ }\bibfield  {title} {\bibinfo {title} {Long range order and
  metastability in two dimensional solids and superfluids. (application of
  dislocation theory)},\ }\href {https://doi.org/10.1088/0022-3719/5/11/002}
  {\bibfield  {journal} {\bibinfo  {journal} {Journal of Physics C: Solid State
  Physics}\ }\textbf {\bibinfo {volume} {5}},\ \bibinfo {pages} {L124}
  (\bibinfo {year} {1972})}\BibitemShut {NoStop}%
\bibitem [{\citenamefont {Kosterlitz}\ and\ \citenamefont
  {Thouless}(1973)}]{Kosterlitz1973}%
  \BibitemOpen
  \bibfield  {author} {\bibinfo {author} {\bibfnamefont {J.~M.}\ \bibnamefont
  {Kosterlitz}}\ and\ \bibinfo {author} {\bibfnamefont {D.~J.}\ \bibnamefont
  {Thouless}},\ }\bibfield  {title} {\bibinfo {title} {Ordering, metastability
  and phase transitions in two-dimensional systems},\ }\href
  {https://doi.org/10.1088/0022-3719/6/7/010} {\bibfield  {journal} {\bibinfo
  {journal} {Journal of Physics C: Solid State Physics}\ }\textbf {\bibinfo
  {volume} {6}},\ \bibinfo {pages} {1181} (\bibinfo {year} {1973})}\BibitemShut
  {NoStop}%
\bibitem [{\citenamefont {Nelson}\ and\ \citenamefont
  {Kosterlitz}(1977)}]{Nelson1977}%
  \BibitemOpen
  \bibfield  {author} {\bibinfo {author} {\bibfnamefont {D.~R.}\ \bibnamefont
  {Nelson}}\ and\ \bibinfo {author} {\bibfnamefont {J.~M.}\ \bibnamefont
  {Kosterlitz}},\ }\bibfield  {title} {\bibinfo {title} {Universal jump in
  superfluid density of 2-dimensional superfluids},\ }\href
  {https://doi.org/DOI 10.1103/PhysRevLett.39.1201} {\bibfield  {journal}
  {\bibinfo  {journal} {Phys. Rev. Lett.}\ }\textbf {\bibinfo {volume} {39}},\
  \bibinfo {pages} {1201} (\bibinfo {year} {1977})}\BibitemShut {NoStop}%
\bibitem [{\citenamefont {Halperin}\ and\ \citenamefont
  {Nelson}(1978)}]{Halperin1978}%
  \BibitemOpen
  \bibfield  {author} {\bibinfo {author} {\bibfnamefont {B.~I.}\ \bibnamefont
  {Halperin}}\ and\ \bibinfo {author} {\bibfnamefont {D.~R.}\ \bibnamefont
  {Nelson}},\ }\bibfield  {title} {\bibinfo {title} {Theory of two-dimensional
  melting},\ }\href {https://doi.org/10.1103/PhysRevLett.41.121} {\bibfield
  {journal} {\bibinfo  {journal} {Phys. Rev. Lett.}\ }\textbf {\bibinfo
  {volume} {41}},\ \bibinfo {pages} {121} (\bibinfo {year} {1978})}\BibitemShut
  {NoStop}%
\bibitem [{\citenamefont {Young}(1979)}]{Young1979}%
  \BibitemOpen
  \bibfield  {author} {\bibinfo {author} {\bibfnamefont {A.~P.}\ \bibnamefont
  {Young}},\ }\bibfield  {title} {\bibinfo {title} {Melting and the vector
  coulomb gas in two dimensions},\ }\href {https://doi.org/DOI
  10.1103/PhysRevB.19.1855} {\bibfield  {journal} {\bibinfo  {journal} {Phys.
  Rev. B}\ }\textbf {\bibinfo {volume} {19}},\ \bibinfo {pages} {1855}
  (\bibinfo {year} {1979})}\BibitemShut {NoStop}%
\bibitem [{\citenamefont {Ritland}(1954)}]{Ritland1954}%
  \BibitemOpen
  \bibfield  {author} {\bibinfo {author} {\bibfnamefont {H.~N.}\ \bibnamefont
  {Ritland}},\ }\bibfield  {title} {\bibinfo {title} {Density phenomena in the
  transformation range of a borosilicate crown glass},\ }\href
  {https://doi.org/https://doi.org/10.1111/j.1151-2916.1954.tb14053.x}
  {\bibfield  {journal} {\bibinfo  {journal} {Journal of the American Ceramic
  Society}\ }\textbf {\bibinfo {volume} {37}},\ \bibinfo {pages} {370}
  (\bibinfo {year} {1954})}\BibitemShut {NoStop}%
\bibitem [{\citenamefont {Vollmayr}\ \emph {et~al.}(1996)\citenamefont
  {Vollmayr}, \citenamefont {Kob},\ and\ \citenamefont
  {Binder}}]{Vollmayr1996}%
  \BibitemOpen
  \bibfield  {author} {\bibinfo {author} {\bibfnamefont {K.}~\bibnamefont
  {Vollmayr}}, \bibinfo {author} {\bibfnamefont {W.}~\bibnamefont {Kob}},\ and\
  \bibinfo {author} {\bibfnamefont {K.}~\bibnamefont {Binder}},\ }\bibfield
  {title} {\bibinfo {title} {Cooling-rate effects in amorphous silica: A
  computer-simulation study},\ }\href
  {https://doi.org/10.1103/PhysRevB.54.15808} {\bibfield  {journal} {\bibinfo
  {journal} {Physical Review B}\ }\textbf {\bibinfo {volume} {54}},\ \bibinfo
  {pages} {15808} (\bibinfo {year} {1996})}\BibitemShut {NoStop}%
\bibitem [{\citenamefont {Angell}\ \emph {et~al.}(2000)\citenamefont {Angell},
  \citenamefont {Ngai}, \citenamefont {McKenna}, \citenamefont {McMillan},\
  and\ \citenamefont {Martin}}]{Angell2000}%
  \BibitemOpen
  \bibfield  {author} {\bibinfo {author} {\bibfnamefont {C.~A.}\ \bibnamefont
  {Angell}}, \bibinfo {author} {\bibfnamefont {K.~L.}\ \bibnamefont {Ngai}},
  \bibinfo {author} {\bibfnamefont {G.~B.}\ \bibnamefont {McKenna}}, \bibinfo
  {author} {\bibfnamefont {P.~F.}\ \bibnamefont {McMillan}},\ and\ \bibinfo
  {author} {\bibfnamefont {S.~W.}\ \bibnamefont {Martin}},\ }\bibfield  {title}
  {\bibinfo {title} {Relaxation in glassforming liquids and amorphous solids},\
  }\href {https://doi.org/10.1063/1.1286035} {\bibfield  {journal} {\bibinfo
  {journal} {Journal of Applied Physics}\ }\textbf {\bibinfo {volume} {88}},\
  \bibinfo {pages} {3113} (\bibinfo {year} {2000})}\BibitemShut {NoStop}%
\bibitem [{\citenamefont {Debenedetti}\ and\ \citenamefont
  {Stillinger}(2001)}]{Debenedetti2001}%
  \BibitemOpen
  \bibfield  {author} {\bibinfo {author} {\bibfnamefont {P.~G.}\ \bibnamefont
  {Debenedetti}}\ and\ \bibinfo {author} {\bibfnamefont {F.~H.}\ \bibnamefont
  {Stillinger}},\ }\bibfield  {title} {\bibinfo {title} {Supercooled liquids
  and the glass transition},\ }\href {https://doi.org/Doi 10.1038/35065704}
  {\bibfield  {journal} {\bibinfo  {journal} {Nat.}\ }\textbf {\bibinfo
  {volume} {410}},\ \bibinfo {pages} {259} (\bibinfo {year}
  {2001})}\BibitemShut {NoStop}%
\bibitem [{\citenamefont {Binder}(1987)}]{Binder1987}%
  \BibitemOpen
  \bibfield  {author} {\bibinfo {author} {\bibfnamefont {K.}~\bibnamefont
  {Binder}},\ }\bibfield  {title} {\bibinfo {title} {Theory of first-order
  phase transitions},\ }\href {http://stacks.iop.org/0034-4885/50/i=7/a=001}
  {\bibfield  {journal} {\bibinfo  {journal} {Rep. Prog. Phys.}\ }\textbf
  {\bibinfo {volume} {50}},\ \bibinfo {pages} {783} (\bibinfo {year}
  {1987})}\BibitemShut {NoStop}%
\bibitem [{\citenamefont {Onsager}(1944)}]{Onsager1944}%
  \BibitemOpen
  \bibfield  {author} {\bibinfo {author} {\bibfnamefont {L.}~\bibnamefont
  {Onsager}},\ }\bibfield  {title} {\bibinfo {title} {Crystal statistics i a
  two-dimensional model with an order-disorder transition},\ }\href
  {https://doi.org/Doi 10.1103/Physrev.65.117} {\bibfield  {journal} {\bibinfo
  {journal} {Phys. Rev.}\ }\textbf {\bibinfo {volume} {65}},\ \bibinfo {pages}
  {117} (\bibinfo {year} {1944})}\BibitemShut {NoStop}%
\bibitem [{\citenamefont {Nelson}\ and\ \citenamefont
  {Halperin}(1979)}]{Nelson1979}%
  \BibitemOpen
  \bibfield  {author} {\bibinfo {author} {\bibfnamefont {D.~R.}\ \bibnamefont
  {Nelson}}\ and\ \bibinfo {author} {\bibfnamefont {B.~I.}\ \bibnamefont
  {Halperin}},\ }\bibfield  {title} {\bibinfo {title} {Dislocation-mediated
  melting in two dimensions},\ }\href {https://doi.org/DOI
  10.1103/PhysRevB.19.2457} {\bibfield  {journal} {\bibinfo  {journal} {Phys.
  Rev. B}\ }\textbf {\bibinfo {volume} {19}},\ \bibinfo {pages} {2457}
  (\bibinfo {year} {1979})}\BibitemShut {NoStop}%
\bibitem [{\citenamefont {Kibble}(1976)}]{Kibble1976}%
  \BibitemOpen
  \bibfield  {author} {\bibinfo {author} {\bibfnamefont {T.~W.~B.}\
  \bibnamefont {Kibble}},\ }\bibfield  {title} {\bibinfo {title} {Topology of
  cosmic domains and strings},\ }\href
  {https://iopscience.iop.org/article/10.1088/0305-4470/9/8/029} {\bibfield
  {journal} {\bibinfo  {journal} {Journal of Physics A: Mathematical and
  General}\ }\textbf {\bibinfo {volume} {9}},\ \bibinfo {pages} {1387}
  (\bibinfo {year} {1976})}\BibitemShut {NoStop}%
\bibitem [{\citenamefont {Kibble}(1980)}]{Kibble1980}%
  \BibitemOpen
  \bibfield  {author} {\bibinfo {author} {\bibfnamefont {T.~W.~B.}\
  \bibnamefont {Kibble}},\ }\bibfield  {title} {\bibinfo {title} {Some
  implications of a cosmological phase transition},\ }\href
  {https://doi.org/http://dx.doi.org/10.1016/0370-1573(80)90091-5} {\bibfield
  {journal} {\bibinfo  {journal} {Physics Reports}\ }\textbf {\bibinfo {volume}
  {67}},\ \bibinfo {pages} {183} (\bibinfo {year} {1980})}\BibitemShut
  {NoStop}%
\bibitem [{\citenamefont {Kibble}(2007)}]{Kibble2007}%
  \BibitemOpen
  \bibfield  {author} {\bibinfo {author} {\bibfnamefont {T.}~\bibnamefont
  {Kibble}},\ }\bibfield  {title} {\bibinfo {title} {Phase-transition dynamics
  in the lab and the universe},\ }\href {https://doi.org/10.1063/1.2784684}
  {\bibfield  {journal} {\bibinfo  {journal} {Physics Today}\ }\textbf
  {\bibinfo {volume} {60}},\ \bibinfo {pages} {47} (\bibinfo {year}
  {2007})}\BibitemShut {NoStop}%
\bibitem [{\citenamefont {Zel'dovic}\ \emph {et~al.}(1974)\citenamefont
  {Zel'dovic}, \citenamefont {Kobzarev},\ and\ \citenamefont
  {L.B.}}]{Zeldovich1974}%
  \BibitemOpen
  \bibfield  {author} {\bibinfo {author} {\bibfnamefont {Y.~B.}\ \bibnamefont
  {Zel'dovic}}, \bibinfo {author} {\bibfnamefont {I.}~\bibnamefont
  {Kobzarev}},\ and\ \bibinfo {author} {\bibfnamefont {O.}~\bibnamefont
  {L.B.}},\ }\bibfield  {title} {\bibinfo {title} {Cosmological consequences of
  a spontaneous breakdown of a discrete symmetry},\ }\href@noop {} {\bibfield
  {journal} {\bibinfo  {journal} {Zh. Eksp. Teor. Fiz.}\ }\textbf {\bibinfo
  {volume} {67}},\ \bibinfo {pages} {3} (\bibinfo {year} {1974})}\BibitemShut
  {NoStop}%
\bibitem [{\citenamefont {Zurek}(1985)}]{Zurek1985}%
  \BibitemOpen
  \bibfield  {author} {\bibinfo {author} {\bibfnamefont {W.~H.}\ \bibnamefont
  {Zurek}},\ }\bibfield  {title} {\bibinfo {title} {Cosmological experiments in
  superfluid helium?},\ }\href {http://dx.doi.org/10.1038/317505a0} {\bibfield
  {journal} {\bibinfo  {journal} {Nature}\ }\textbf {\bibinfo {volume} {317}},\
  \bibinfo {pages} {505} (\bibinfo {year} {1985})}\BibitemShut {NoStop}%
\bibitem [{\citenamefont {Zurek}(1993)}]{Zurek1993}%
  \BibitemOpen
  \bibfield  {author} {\bibinfo {author} {\bibfnamefont {W.~H.}\ \bibnamefont
  {Zurek}},\ }\bibfield  {title} {\bibinfo {title} {Cosmic strings in
  laboratory superfluids and the topological remnants of other phase
  transitions},\ }\href
  {http://www.actaphys.uj.edu.pl/vol24/abs/v24p1301.htm?series=reg&vol=24&page=1301}
  {\bibfield  {journal} {\bibinfo  {journal} {Acta Phys Pol B}\ }\textbf
  {\bibinfo {volume} {24}} (\bibinfo {year} {1993})}\BibitemShut {NoStop}%
\bibitem [{\citenamefont {Zurek}(1996)}]{Zurek1996}%
  \BibitemOpen
  \bibfield  {author} {\bibinfo {author} {\bibfnamefont {W.~H.}\ \bibnamefont
  {Zurek}},\ }\bibfield  {title} {\bibinfo {title} {Cosmological experiments in
  condensed matter systems},\ }\href
  {https://doi.org/https://doi.org/10.1016/S0370-1573(96)00009-9} {\bibfield
  {journal} {\bibinfo  {journal} {Physics Reports}\ }\textbf {\bibinfo {volume}
  {276}},\ \bibinfo {pages} {177} (\bibinfo {year} {1996})}\BibitemShut
  {NoStop}%
\bibitem [{\citenamefont {Chuang}\ \emph {et~al.}(1991)\citenamefont {Chuang},
  \citenamefont {Durrer}, \citenamefont {Turok},\ and\ \citenamefont
  {Yurke}}]{Chuang1991}%
  \BibitemOpen
  \bibfield  {author} {\bibinfo {author} {\bibfnamefont {I.}~\bibnamefont
  {Chuang}}, \bibinfo {author} {\bibfnamefont {R.}~\bibnamefont {Durrer}},
  \bibinfo {author} {\bibfnamefont {N.}~\bibnamefont {Turok}},\ and\ \bibinfo
  {author} {\bibfnamefont {B.}~\bibnamefont {Yurke}},\ }\bibfield  {title}
  {\bibinfo {title} {Cosmology in the laboratory: Defect dynamics in liquid
  crystals},\ }\href {https://doi.org/10.1126/science.251.4999.1336} {\bibfield
   {journal} {\bibinfo  {journal} {Science}\ }\textbf {\bibinfo {volume}
  {251}},\ \bibinfo {pages} {1336} (\bibinfo {year} {1991})}\BibitemShut
  {NoStop}%
\bibitem [{\citenamefont {Bauerle}\ \emph {et~al.}(1996)\citenamefont
  {Bauerle}, \citenamefont {Bunkov}, \citenamefont {Fisher}, \citenamefont
  {Godfrin},\ and\ \citenamefont {Pickett}}]{Bauerle1996}%
  \BibitemOpen
  \bibfield  {author} {\bibinfo {author} {\bibfnamefont {C.}~\bibnamefont
  {Bauerle}}, \bibinfo {author} {\bibfnamefont {Y.~M.}\ \bibnamefont {Bunkov}},
  \bibinfo {author} {\bibfnamefont {S.~N.}\ \bibnamefont {Fisher}}, \bibinfo
  {author} {\bibfnamefont {H.}~\bibnamefont {Godfrin}},\ and\ \bibinfo {author}
  {\bibfnamefont {G.~R.}\ \bibnamefont {Pickett}},\ }\bibfield  {title}
  {\bibinfo {title} {Laboratory simulation of cosmic string formation in the
  early universe using superfluid 3he},\ }\href
  {http://dx.doi.org/10.1038/382332a0} {\bibfield  {journal} {\bibinfo
  {journal} {Nature}\ }\textbf {\bibinfo {volume} {382}},\ \bibinfo {pages}
  {332} (\bibinfo {year} {1996})}\BibitemShut {NoStop}%
\bibitem [{\citenamefont {Carmi}\ \emph {et~al.}(2000)\citenamefont {Carmi},
  \citenamefont {Polturak},\ and\ \citenamefont {Koren}}]{Carmi2000}%
  \BibitemOpen
  \bibfield  {author} {\bibinfo {author} {\bibfnamefont {R.}~\bibnamefont
  {Carmi}}, \bibinfo {author} {\bibfnamefont {E.}~\bibnamefont {Polturak}},\
  and\ \bibinfo {author} {\bibfnamefont {G.}~\bibnamefont {Koren}},\ }\bibfield
   {title} {\bibinfo {title} {Observation of spontaneous flux generation in a
  multi-josephson-junction loop},\ }\href
  {https://link.aps.org/doi/10.1103/PhysRevLett.84.4966} {\bibfield  {journal}
  {\bibinfo  {journal} {Physical Review Letters}\ }\textbf {\bibinfo {volume}
  {84}},\ \bibinfo {pages} {4966} (\bibinfo {year} {2000})}\BibitemShut
  {NoStop}%
\bibitem [{\citenamefont {Miranda}\ \emph {et~al.}(2013)\citenamefont
  {Miranda}, \citenamefont {Burguete}, \citenamefont {Mancini},\ and\
  \citenamefont {González-Viñas}}]{Miranda2013}%
  \BibitemOpen
  \bibfield  {author} {\bibinfo {author} {\bibfnamefont {M.~A.}\ \bibnamefont
  {Miranda}}, \bibinfo {author} {\bibfnamefont {J.}~\bibnamefont {Burguete}},
  \bibinfo {author} {\bibfnamefont {H.}~\bibnamefont {Mancini}},\ and\ \bibinfo
  {author} {\bibfnamefont {W.}~\bibnamefont {González-Viñas}},\ }\bibfield
  {title} {\bibinfo {title} {Frozen dynamics and synchronization through a
  secondary symmetry-breaking bifurcation},\ }\href
  {https://doi.org/10.1103/PhysRevE.87.032902} {\bibfield  {journal} {\bibinfo
  {journal} {Physical Review E}\ }\textbf {\bibinfo {volume} {87}},\ \bibinfo
  {pages} {032902} (\bibinfo {year} {2013})}\BibitemShut {NoStop}%
\bibitem [{\citenamefont {Chae}\ \emph {et~al.}(2012)\citenamefont {Chae},
  \citenamefont {Lee}, \citenamefont {Horibe}, \citenamefont {Tanimura},
  \citenamefont {Mori}, \citenamefont {Gao}, \citenamefont {Carr},\ and\
  \citenamefont {Cheong}}]{Chae2012}%
  \BibitemOpen
  \bibfield  {author} {\bibinfo {author} {\bibfnamefont {S.~C.}\ \bibnamefont
  {Chae}}, \bibinfo {author} {\bibfnamefont {N.}~\bibnamefont {Lee}}, \bibinfo
  {author} {\bibfnamefont {Y.}~\bibnamefont {Horibe}}, \bibinfo {author}
  {\bibfnamefont {M.}~\bibnamefont {Tanimura}}, \bibinfo {author}
  {\bibfnamefont {S.}~\bibnamefont {Mori}}, \bibinfo {author} {\bibfnamefont
  {B.}~\bibnamefont {Gao}}, \bibinfo {author} {\bibfnamefont {S.}~\bibnamefont
  {Carr}},\ and\ \bibinfo {author} {\bibfnamefont {S.~W.}\ \bibnamefont
  {Cheong}},\ }\bibfield  {title} {\bibinfo {title} {Direct observation of the
  proliferation of ferroelectric loop domains and vortex-antivortex pairs},\
  }\href {https://doi.org/10.1103/PhysRevLett.108.167603} {\bibfield  {journal}
  {\bibinfo  {journal} {Physical Review Letters}\ }\textbf {\bibinfo {volume}
  {108}},\ \bibinfo {pages} {167603} (\bibinfo {year} {2012})}\BibitemShut
  {NoStop}%
\bibitem [{\citenamefont {Xu}\ \emph {et~al.}(2014)\citenamefont {Xu},
  \citenamefont {Han}, \citenamefont {Sun}, \citenamefont {Xu}, \citenamefont
  {Tang}, \citenamefont {Li},\ and\ \citenamefont {Guo}}]{Xu2014}%
  \BibitemOpen
  \bibfield  {author} {\bibinfo {author} {\bibfnamefont {X.-Y.}\ \bibnamefont
  {Xu}}, \bibinfo {author} {\bibfnamefont {Y.-J.}\ \bibnamefont {Han}},
  \bibinfo {author} {\bibfnamefont {K.}~\bibnamefont {Sun}}, \bibinfo {author}
  {\bibfnamefont {J.-S.}\ \bibnamefont {Xu}}, \bibinfo {author} {\bibfnamefont
  {J.-S.}\ \bibnamefont {Tang}}, \bibinfo {author} {\bibfnamefont {C.-F.}\
  \bibnamefont {Li}},\ and\ \bibinfo {author} {\bibfnamefont {G.-C.}\
  \bibnamefont {Guo}},\ }\bibfield  {title} {\bibinfo {title} {Quantum
  simulation of landau-zener model dynamics supporting the kibble-zurek
  mechanism},\ }\href {https://doi.org/10.1103/PhysRevLett.112.035701}
  {\bibfield  {journal} {\bibinfo  {journal} {Physical Review Letters}\
  }\textbf {\bibinfo {volume} {112}},\ \bibinfo {pages} {035701} (\bibinfo
  {year} {2014})}\BibitemShut {NoStop}%
\bibitem [{\citenamefont {Ulm}\ \emph {et~al.}(2013)\citenamefont {Ulm},
  \citenamefont {Roßnagel}, \citenamefont {Jacob}, \citenamefont {Degünther},
  \citenamefont {Dawkins}, \citenamefont {Poschinger}, \citenamefont
  {Nigmatullin}, \citenamefont {Retzker}, \citenamefont {Plenio}, \citenamefont
  {Schmidt-Kaler},\ and\ \citenamefont {Singer}}]{Ulm2013}%
  \BibitemOpen
  \bibfield  {author} {\bibinfo {author} {\bibfnamefont {S.}~\bibnamefont
  {Ulm}}, \bibinfo {author} {\bibfnamefont {J.}~\bibnamefont {Roßnagel}},
  \bibinfo {author} {\bibfnamefont {G.}~\bibnamefont {Jacob}}, \bibinfo
  {author} {\bibfnamefont {C.}~\bibnamefont {Degünther}}, \bibinfo {author}
  {\bibfnamefont {S.~T.}\ \bibnamefont {Dawkins}}, \bibinfo {author}
  {\bibfnamefont {U.~G.}\ \bibnamefont {Poschinger}}, \bibinfo {author}
  {\bibfnamefont {R.}~\bibnamefont {Nigmatullin}}, \bibinfo {author}
  {\bibfnamefont {A.}~\bibnamefont {Retzker}}, \bibinfo {author} {\bibfnamefont
  {M.~B.}\ \bibnamefont {Plenio}}, \bibinfo {author} {\bibfnamefont
  {F.}~\bibnamefont {Schmidt-Kaler}},\ and\ \bibinfo {author} {\bibfnamefont
  {K.}~\bibnamefont {Singer}},\ }\bibfield  {title} {\bibinfo {title}
  {Observation of the kibble–zurek scaling law for defect formation in ion
  crystals},\ }\href {https://doi.org/10.1038/ncomms3290} {\bibfield  {journal}
  {\bibinfo  {journal} {Nature Communications}\ }\textbf {\bibinfo {volume}
  {4}},\ \bibinfo {pages} {2290} (\bibinfo {year} {2013})}\BibitemShut
  {NoStop}%
\bibitem [{\citenamefont {Pyka}\ \emph {et~al.}(2013)\citenamefont {Pyka},
  \citenamefont {Keller}, \citenamefont {Partner}, \citenamefont {Nigmatullin},
  \citenamefont {Burgermeister}, \citenamefont {Meier}, \citenamefont
  {Kuhlmann}, \citenamefont {Retzker}, \citenamefont {Plenio}, \citenamefont
  {Zurek}, \citenamefont {del Campo},\ and\ \citenamefont
  {Mehlstäubler}}]{Pyka2013}%
  \BibitemOpen
  \bibfield  {author} {\bibinfo {author} {\bibfnamefont {K.}~\bibnamefont
  {Pyka}}, \bibinfo {author} {\bibfnamefont {J.}~\bibnamefont {Keller}},
  \bibinfo {author} {\bibfnamefont {H.~L.}\ \bibnamefont {Partner}}, \bibinfo
  {author} {\bibfnamefont {R.}~\bibnamefont {Nigmatullin}}, \bibinfo {author}
  {\bibfnamefont {T.}~\bibnamefont {Burgermeister}}, \bibinfo {author}
  {\bibfnamefont {D.~M.}\ \bibnamefont {Meier}}, \bibinfo {author}
  {\bibfnamefont {K.}~\bibnamefont {Kuhlmann}}, \bibinfo {author}
  {\bibfnamefont {A.}~\bibnamefont {Retzker}}, \bibinfo {author} {\bibfnamefont
  {M.~B.}\ \bibnamefont {Plenio}}, \bibinfo {author} {\bibfnamefont {W.~H.}\
  \bibnamefont {Zurek}}, \bibinfo {author} {\bibfnamefont {A.}~\bibnamefont
  {del Campo}},\ and\ \bibinfo {author} {\bibfnamefont {T.~E.}\ \bibnamefont
  {Mehlstäubler}},\ }\bibfield  {title} {\bibinfo {title} {Topological defect
  formation and spontaneous symmetry breaking in ion coulomb crystals},\ }\href
  {https://doi.org/10.1038/ncomms3291
  https://www.nature.com/articles/ncomms3291#supplementary-information}
  {\bibfield  {journal} {\bibinfo  {journal} {Nature Communications}\ }\textbf
  {\bibinfo {volume} {4}},\ \bibinfo {pages} {2291} (\bibinfo {year}
  {2013})}\BibitemShut {NoStop}%
\bibitem [{\citenamefont {Lamporesi}\ \emph {et~al.}(2013)\citenamefont
  {Lamporesi}, \citenamefont {Donadello}, \citenamefont {Serafini},
  \citenamefont {Dalfovo},\ and\ \citenamefont {Ferrari}}]{Lamporesi2013}%
  \BibitemOpen
  \bibfield  {author} {\bibinfo {author} {\bibfnamefont {G.}~\bibnamefont
  {Lamporesi}}, \bibinfo {author} {\bibfnamefont {S.}~\bibnamefont
  {Donadello}}, \bibinfo {author} {\bibfnamefont {S.}~\bibnamefont {Serafini}},
  \bibinfo {author} {\bibfnamefont {F.}~\bibnamefont {Dalfovo}},\ and\ \bibinfo
  {author} {\bibfnamefont {G.}~\bibnamefont {Ferrari}},\ }\bibfield  {title}
  {\bibinfo {title} {Spontaneous creation of kibble–zurek solitons in a
  bose–einstein condensate},\ }\href {https://doi.org/10.1038/nphys2734}
  {\bibfield  {journal} {\bibinfo  {journal} {Nature Physics}\ }\textbf
  {\bibinfo {volume} {9}},\ \bibinfo {pages} {656} (\bibinfo {year}
  {2013})}\BibitemShut {NoStop}%
\bibitem [{\citenamefont {del Campo}\ and\ \citenamefont
  {Zurek}(2014)}]{Campo2014}%
  \BibitemOpen
  \bibfield  {author} {\bibinfo {author} {\bibfnamefont {A.}~\bibnamefont {del
  Campo}}\ and\ \bibinfo {author} {\bibfnamefont {W.~H.}\ \bibnamefont
  {Zurek}},\ }\bibfield  {title} {\bibinfo {title} {Universality of phase
  transition dynamics: Topological defects from symmetry breaking},\ }\href
  {https://doi.org/doi:10.1142/S0217751X1430018X} {\bibfield  {journal}
  {\bibinfo  {journal} {International Journal of Modern Physics A}\ }\textbf
  {\bibinfo {volume} {29}},\ \bibinfo {pages} {1430018} (\bibinfo {year}
  {2014})}\BibitemShut {NoStop}%
\bibitem [{\citenamefont {Keesling}\ \emph {et~al.}(2019)\citenamefont
  {Keesling}, \citenamefont {Omran}, \citenamefont {Levine}, \citenamefont
  {Bernien}, \citenamefont {Pichler}, \citenamefont {Choi}, \citenamefont
  {Samajdar}, \citenamefont {Schwartz}, \citenamefont {Silvi}, \citenamefont
  {Sachdev}, \citenamefont {Zoller}, \citenamefont {Endres}, \citenamefont
  {Greiner}, \citenamefont {Vuletić},\ and\ \citenamefont
  {Lukin}}]{Keesling2019}%
  \BibitemOpen
  \bibfield  {author} {\bibinfo {author} {\bibfnamefont {A.}~\bibnamefont
  {Keesling}}, \bibinfo {author} {\bibfnamefont {A.}~\bibnamefont {Omran}},
  \bibinfo {author} {\bibfnamefont {H.}~\bibnamefont {Levine}}, \bibinfo
  {author} {\bibfnamefont {H.}~\bibnamefont {Bernien}}, \bibinfo {author}
  {\bibfnamefont {H.}~\bibnamefont {Pichler}}, \bibinfo {author} {\bibfnamefont
  {S.}~\bibnamefont {Choi}}, \bibinfo {author} {\bibfnamefont {R.}~\bibnamefont
  {Samajdar}}, \bibinfo {author} {\bibfnamefont {S.}~\bibnamefont {Schwartz}},
  \bibinfo {author} {\bibfnamefont {P.}~\bibnamefont {Silvi}}, \bibinfo
  {author} {\bibfnamefont {S.}~\bibnamefont {Sachdev}}, \bibinfo {author}
  {\bibfnamefont {P.}~\bibnamefont {Zoller}}, \bibinfo {author} {\bibfnamefont
  {M.}~\bibnamefont {Endres}}, \bibinfo {author} {\bibfnamefont
  {M.}~\bibnamefont {Greiner}}, \bibinfo {author} {\bibfnamefont
  {V.}~\bibnamefont {Vuletić}},\ and\ \bibinfo {author} {\bibfnamefont
  {M.~D.}\ \bibnamefont {Lukin}},\ }\bibfield  {title} {\bibinfo {title}
  {Quantum kibble–zurek mechanism and critical dynamics on a programmable
  rydberg simulator},\ }\href {https://doi.org/10.1038/s41586-019-1070-1}
  {\bibfield  {journal} {\bibinfo  {journal} {Nature}\ }\textbf {\bibinfo
  {volume} {568}},\ \bibinfo {pages} {207} (\bibinfo {year}
  {2019})}\BibitemShut {NoStop}%
\bibitem [{\citenamefont {Ebadi}\ \emph {et~al.}(2021)\citenamefont {Ebadi},
  \citenamefont {Wang}, \citenamefont {Levine}, \citenamefont {Keesling},
  \citenamefont {Semeghini}, \citenamefont {Omran}, \citenamefont {Bluvstein},
  \citenamefont {Samajdar}, \citenamefont {Pichler}, \citenamefont {Ho},
  \citenamefont {Choi}, \citenamefont {Sachdev}, \citenamefont {Greiner},
  \citenamefont {Vuletić},\ and\ \citenamefont {Lukin}}]{Ebadi2021}%
  \BibitemOpen
  \bibfield  {author} {\bibinfo {author} {\bibfnamefont {S.}~\bibnamefont
  {Ebadi}}, \bibinfo {author} {\bibfnamefont {T.~T.}\ \bibnamefont {Wang}},
  \bibinfo {author} {\bibfnamefont {H.}~\bibnamefont {Levine}}, \bibinfo
  {author} {\bibfnamefont {A.}~\bibnamefont {Keesling}}, \bibinfo {author}
  {\bibfnamefont {G.}~\bibnamefont {Semeghini}}, \bibinfo {author}
  {\bibfnamefont {A.}~\bibnamefont {Omran}}, \bibinfo {author} {\bibfnamefont
  {D.}~\bibnamefont {Bluvstein}}, \bibinfo {author} {\bibfnamefont
  {R.}~\bibnamefont {Samajdar}}, \bibinfo {author} {\bibfnamefont
  {H.}~\bibnamefont {Pichler}}, \bibinfo {author} {\bibfnamefont {W.~W.}\
  \bibnamefont {Ho}}, \bibinfo {author} {\bibfnamefont {S.}~\bibnamefont
  {Choi}}, \bibinfo {author} {\bibfnamefont {S.}~\bibnamefont {Sachdev}},
  \bibinfo {author} {\bibfnamefont {M.}~\bibnamefont {Greiner}}, \bibinfo
  {author} {\bibfnamefont {V.}~\bibnamefont {Vuletić}},\ and\ \bibinfo
  {author} {\bibfnamefont {M.~D.}\ \bibnamefont {Lukin}},\ }\bibfield  {title}
  {\bibinfo {title} {Quantum phases of matter on a 256-atom programmable
  quantum simulator},\ }\href {https://doi.org/10.1038/s41586-021-03582-4}
  {\bibfield  {journal} {\bibinfo  {journal} {Nature}\ }\textbf {\bibinfo
  {volume} {595}},\ \bibinfo {pages} {227} (\bibinfo {year}
  {2021})}\BibitemShut {NoStop}%
\bibitem [{\citenamefont {Schmitt}\ \emph {et~al.}(2022)\citenamefont
  {Schmitt}, \citenamefont {Rams}, \citenamefont {Dziarmaga}, \citenamefont
  {Heyl},\ and\ \citenamefont {Zurek}}]{Schmitt2022}%
  \BibitemOpen
  \bibfield  {author} {\bibinfo {author} {\bibfnamefont {M.}~\bibnamefont
  {Schmitt}}, \bibinfo {author} {\bibfnamefont {M.~M.}\ \bibnamefont {Rams}},
  \bibinfo {author} {\bibfnamefont {J.}~\bibnamefont {Dziarmaga}}, \bibinfo
  {author} {\bibfnamefont {M.}~\bibnamefont {Heyl}},\ and\ \bibinfo {author}
  {\bibfnamefont {W.~H.}\ \bibnamefont {Zurek}},\ }\bibfield  {title} {\bibinfo
  {title} {Quantum phase transition dynamics in the two-dimensional
  transverse-field ising model},\ }\href
  {https://doi.org/doi:10.1126/sciadv.abl6850} {\bibfield  {journal} {\bibinfo
  {journal} {Science Advances}\ }\textbf {\bibinfo {volume} {8}},\ \bibinfo
  {pages} {eabl6850} (\bibinfo {year} {2022})}\BibitemShut {NoStop}%
\bibitem [{\citenamefont {del Campo}(2018)}]{Campo2018}%
  \BibitemOpen
  \bibfield  {author} {\bibinfo {author} {\bibfnamefont {A.}~\bibnamefont {del
  Campo}},\ }\bibfield  {title} {\bibinfo {title} {Universal statistics of
  topological defects formed in a quantum phase transition},\ }\href
  {https://doi.org/10.1103/PhysRevLett.121.200601} {\bibfield  {journal}
  {\bibinfo  {journal} {Physical Review Letters}\ }\textbf {\bibinfo {volume}
  {121}},\ \bibinfo {pages} {200601} (\bibinfo {year} {2018})}\BibitemShut
  {NoStop}%
\bibitem [{\citenamefont {Zeng}\ \emph {et~al.}(2023)\citenamefont {Zeng},
  \citenamefont {Xia},\ and\ \citenamefont {del Campo}}]{Zeng2023}%
  \BibitemOpen
  \bibfield  {author} {\bibinfo {author} {\bibfnamefont {H.-B.}\ \bibnamefont
  {Zeng}}, \bibinfo {author} {\bibfnamefont {C.-Y.}\ \bibnamefont {Xia}},\ and\
  \bibinfo {author} {\bibfnamefont {A.}~\bibnamefont {del Campo}},\ }\bibfield
  {title} {\bibinfo {title} {Universal breakdown of kibble-zurek scaling in
  fast quenches across a phase transition},\ }\href
  {https://doi.org/10.1103/PhysRevLett.130.060402} {\bibfield  {journal}
  {\bibinfo  {journal} {Physical Review Letters}\ }\textbf {\bibinfo {volume}
  {130}},\ \bibinfo {pages} {060402} (\bibinfo {year} {2023})}\BibitemShut
  {NoStop}%
\bibitem [{\citenamefont {Deutschländer}\ \emph {et~al.}(2015)\citenamefont
  {Deutschländer}, \citenamefont {Dillmann}, \citenamefont {Maret},\ and\
  \citenamefont {Keim}}]{Deutschlaender2015}%
  \BibitemOpen
  \bibfield  {author} {\bibinfo {author} {\bibfnamefont {S.}~\bibnamefont
  {Deutschländer}}, \bibinfo {author} {\bibfnamefont {P.}~\bibnamefont
  {Dillmann}}, \bibinfo {author} {\bibfnamefont {G.}~\bibnamefont {Maret}},\
  and\ \bibinfo {author} {\bibfnamefont {P.}~\bibnamefont {Keim}},\ }\bibfield
  {title} {\bibinfo {title} {Kibble–zurek mechanism in colloidal
  monolayers},\ }\href
  {https://doi.org/https://doi.org/10.1073/pnas.1500763112} {\bibfield
  {journal} {\bibinfo  {journal} {Proc. Nat. Acad. Sci.}\ }\textbf {\bibinfo
  {volume} {112}},\ \bibinfo {pages} {6925} (\bibinfo {year}
  {2015})}\BibitemShut {NoStop}%
\bibitem [{\citenamefont {Ebert}\ \emph {et~al.}(2009)\citenamefont {Ebert},
  \citenamefont {Dillmann}, \citenamefont {Maret},\ and\ \citenamefont
  {Keim}}]{ebert2009a}%
  \BibitemOpen
  \bibfield  {author} {\bibinfo {author} {\bibfnamefont {F.}~\bibnamefont
  {Ebert}}, \bibinfo {author} {\bibfnamefont {P.}~\bibnamefont {Dillmann}},
  \bibinfo {author} {\bibfnamefont {G.}~\bibnamefont {Maret}},\ and\ \bibinfo
  {author} {\bibfnamefont {P.}~\bibnamefont {Keim}},\ }\bibfield  {title}
  {\bibinfo {title} {The experimental realization of a two-dimensional
  colloidal model system},\ }\href
  {https://doi.org/https://doi.org/10.1063/1.3188948} {\bibfield  {journal}
  {\bibinfo  {journal} {Rev. Sci. Inst.}\ }\textbf {\bibinfo {volume} {80}},\
  \bibinfo {pages} {083902} (\bibinfo {year} {2009})}\BibitemShut {NoStop}%
\bibitem [{\citenamefont {Keim}\ \emph {et~al.}(2007)\citenamefont {Keim},
  \citenamefont {Maret},\ and\ \citenamefont {von Grünberg}}]{Keim2007}%
  \BibitemOpen
  \bibfield  {author} {\bibinfo {author} {\bibfnamefont {P.}~\bibnamefont
  {Keim}}, \bibinfo {author} {\bibfnamefont {G.}~\bibnamefont {Maret}},\ and\
  \bibinfo {author} {\bibfnamefont {H.~H.}\ \bibnamefont {von Grünberg}},\
  }\bibfield  {title} {\bibinfo {title} {Frank's constant in the hexatic
  phase},\ }\href {https://doi.org/https://doi.org/10.1103/PhysRevE.75.031402}
  {\bibfield  {journal} {\bibinfo  {journal} {Phys. Rev. E.}\ }\textbf
  {\bibinfo {volume} {75}},\ \bibinfo {pages} {031402} (\bibinfo {year}
  {2007})}\BibitemShut {NoStop}%
\bibitem [{\citenamefont {Gasser}\ \emph {et~al.}(2010)\citenamefont {Gasser},
  \citenamefont {Eisenmann}, \citenamefont {Maret},\ and\ \citenamefont
  {Keim}}]{Gasser2010}%
  \BibitemOpen
  \bibfield  {author} {\bibinfo {author} {\bibfnamefont {U.}~\bibnamefont
  {Gasser}}, \bibinfo {author} {\bibfnamefont {C.}~\bibnamefont {Eisenmann}},
  \bibinfo {author} {\bibfnamefont {G.}~\bibnamefont {Maret}},\ and\ \bibinfo
  {author} {\bibfnamefont {P.}~\bibnamefont {Keim}},\ }\bibfield  {title}
  {\bibinfo {title} {Melting of crystals in two dimensions},\ }\href
  {https://doi.org/https://doi.org/10.1002/cphc.200900755} {\bibfield
  {journal} {\bibinfo  {journal} {ChemPhysChem}\ }\textbf {\bibinfo {volume}
  {11}},\ \bibinfo {pages} {963} (\bibinfo {year} {2010})}\BibitemShut
  {NoStop}%
\bibitem [{\citenamefont {Kosterlitz}(2017)}]{Kosterlitz2017}%
  \BibitemOpen
  \bibfield  {author} {\bibinfo {author} {\bibfnamefont {J.~M.}\ \bibnamefont
  {Kosterlitz}},\ }\bibfield  {title} {\bibinfo {title} {Nobel lecture:
  Topological defects and phase transitions},\ }\href
  {https://link.aps.org/doi/10.1103/RevModPhys.89.040501} {\bibfield  {journal}
  {\bibinfo  {journal} {Reviews of Modern Physics}\ }\textbf {\bibinfo {volume}
  {89}},\ \bibinfo {pages} {040501} (\bibinfo {year} {2017})}\BibitemShut
  {NoStop}%
\bibitem [{\citenamefont {Illing}\ \emph {et~al.}(2017)\citenamefont {Illing},
  \citenamefont {Fritschi}, \citenamefont {Kaiser}, \citenamefont {Klix},
  \citenamefont {Maret},\ and\ \citenamefont {Keim}}]{Illing2017}%
  \BibitemOpen
  \bibfield  {author} {\bibinfo {author} {\bibfnamefont {B.}~\bibnamefont
  {Illing}}, \bibinfo {author} {\bibfnamefont {S.}~\bibnamefont {Fritschi}},
  \bibinfo {author} {\bibfnamefont {H.}~\bibnamefont {Kaiser}}, \bibinfo
  {author} {\bibfnamefont {C.~L.}\ \bibnamefont {Klix}}, \bibinfo {author}
  {\bibfnamefont {G.}~\bibnamefont {Maret}},\ and\ \bibinfo {author}
  {\bibfnamefont {P.}~\bibnamefont {Keim}},\ }\bibfield  {title} {\bibinfo
  {title} {Mermin–wagner fluctuations in 2d amorphous solids},\ }\href
  {https://doi.org/https://doi.org/10.1073/pnas.1612964114} {\bibfield
  {journal} {\bibinfo  {journal} {Proc. Nat. Acad. Sci.}\ }\textbf {\bibinfo
  {volume} {114}},\ \bibinfo {pages} {1856} (\bibinfo {year}
  {2017})}\BibitemShut {NoStop}%
\bibitem [{Note1()}]{Note1}%
  \BibitemOpen
  \bibinfo {note} {Dynabeads, 4.5 $\mu $m, http://www.dynal.no}\BibitemShut
  {NoStop}%
\bibitem [{Note2()}]{Note2}%
  \BibitemOpen
  \bibinfo {note} {Earlier work \cite {Keim2007} reports lower values due to a
  lower magnetic susceptibility of the colloids, measured by SQUID. Meanwhile,
  it is more precise to determine the susceptibility in comparison of the pair
  correlation function in the fluid phase with computer simulations, leading to
  the given transition temperatures.}\BibitemShut {Stop}%
\bibitem [{\citenamefont {Mermin}\ and\ \citenamefont
  {Wagner}(1966)}]{Mermin1966}%
  \BibitemOpen
  \bibfield  {author} {\bibinfo {author} {\bibfnamefont {N.~D.}\ \bibnamefont
  {Mermin}}\ and\ \bibinfo {author} {\bibfnamefont {H.}~\bibnamefont
  {Wagner}},\ }\bibfield  {title} {\bibinfo {title} {Absence of ferromagnetism
  or antiferromagnetism in one- or two-dimensional isotropic heisenberg
  models},\ }\href {https://doi.org/DOI 10.1103/PhysRevLett.17.1133} {\bibfield
   {journal} {\bibinfo  {journal} {Phys. Rev. Lett.}\ }\textbf {\bibinfo
  {volume} {17}},\ \bibinfo {pages} {1133} (\bibinfo {year}
  {1966})}\BibitemShut {NoStop}%
\bibitem [{\citenamefont {Hohenberg}(1967)}]{Hohenberg1967}%
  \BibitemOpen
  \bibfield  {author} {\bibinfo {author} {\bibfnamefont {P.}~\bibnamefont
  {Hohenberg}},\ }\bibfield  {title} {\bibinfo {title} {Existence of long-range
  order in 1 and 2 dimensions},\ }\href
  {https://doi.org/10.1103/PhysRev.158.383} {\bibfield  {journal} {\bibinfo
  {journal} {Phys. Rev.}\ }\textbf {\bibinfo {volume} {158}},\ \bibinfo {pages}
  {383} (\bibinfo {year} {1967})}\BibitemShut {NoStop}%
\bibitem [{\citenamefont {Mermin}(1968)}]{Mermin1968}%
  \BibitemOpen
  \bibfield  {author} {\bibinfo {author} {\bibfnamefont {N.~D.}\ \bibnamefont
  {Mermin}},\ }\bibfield  {title} {\bibinfo {title} {Crystalline order in two
  dimensions},\ }\href {https://doi.org/10.1103/PhysRev.176.250} {\bibfield
  {journal} {\bibinfo  {journal} {Phys. Rev.}\ }\textbf {\bibinfo {volume}
  {176}},\ \bibinfo {pages} {250} (\bibinfo {year} {1968})}\BibitemShut
  {NoStop}%
\bibitem [{\citenamefont {Dillmann}\ \emph {et~al.}(2013)\citenamefont
  {Dillmann}, \citenamefont {Maret},\ and\ \citenamefont
  {Keim}}]{Dillmann2013}%
  \BibitemOpen
  \bibfield  {author} {\bibinfo {author} {\bibfnamefont {P.}~\bibnamefont
  {Dillmann}}, \bibinfo {author} {\bibfnamefont {G.}~\bibnamefont {Maret}},\
  and\ \bibinfo {author} {\bibfnamefont {P.}~\bibnamefont {Keim}},\ }\bibfield
  {title} {\bibinfo {title} {Two-dimensional colloidal systems in
  time-dependent magnetic fields},\ }\href
  {https://doi.org/http://dx.doi.org/10.1140/epjst/e2013-02068-9} {\bibfield
  {journal} {\bibinfo  {journal} {Eur. Phys. J. Special Topics}\ }\textbf
  {\bibinfo {volume} {222}},\ \bibinfo {pages} {2941} (\bibinfo {year}
  {2013})}\BibitemShut {NoStop}%
\bibitem [{\citenamefont {Biroli}\ \emph {et~al.}(2010)\citenamefont {Biroli},
  \citenamefont {Cugliandolo},\ and\ \citenamefont {Sicilia}}]{Biroli2010}%
  \BibitemOpen
  \bibfield  {author} {\bibinfo {author} {\bibfnamefont {G.}~\bibnamefont
  {Biroli}}, \bibinfo {author} {\bibfnamefont {L.~F.}\ \bibnamefont
  {Cugliandolo}},\ and\ \bibinfo {author} {\bibfnamefont {A.}~\bibnamefont
  {Sicilia}},\ }\bibfield  {title} {\bibinfo {title} {Kibble-zurek mechanism
  and infinitely slow annealing through critical points},\ }\href
  {https://link.aps.org/doi/10.1103/PhysRevE.81.050101} {\bibfield  {journal}
  {\bibinfo  {journal} {Physical Review E}\ }\textbf {\bibinfo {volume} {81}},\
  \bibinfo {pages} {050101} (\bibinfo {year} {2010})}\BibitemShut {NoStop}%
\bibitem [{\citenamefont {Asja}\ and\ \citenamefont
  {Leticia}(2011)}]{Jelic2011}%
  \BibitemOpen
  \bibfield  {author} {\bibinfo {author} {\bibfnamefont {J.}~\bibnamefont
  {Asja}}\ and\ \bibinfo {author} {\bibfnamefont {F.~C.}\ \bibnamefont
  {Leticia}},\ }\bibfield  {title} {\bibinfo {title} {Quench dynamics of the 2d
  xy model},\ }\href {http://stacks.iop.org/1742-5468/2011/i=02/a=P02032}
  {\bibfield  {journal} {\bibinfo  {journal} {Journal of Statistical Mechanics:
  Theory and Experiment}\ }\textbf {\bibinfo {volume} {2011}},\ \bibinfo
  {pages} {P02032} (\bibinfo {year} {2011})}\BibitemShut {NoStop}%
\bibitem [{\citenamefont {Hartmann}\ \emph {et~al.}(2010)\citenamefont
  {Hartmann}, \citenamefont {Douglass}, \citenamefont {Reyes}, \citenamefont
  {Matthews}, \citenamefont {Hyde}, \citenamefont {Kovacs},\ and\ \citenamefont
  {Donko}}]{Hartmann2010}%
  \BibitemOpen
  \bibfield  {author} {\bibinfo {author} {\bibfnamefont {P.}~\bibnamefont
  {Hartmann}}, \bibinfo {author} {\bibfnamefont {A.}~\bibnamefont {Douglass}},
  \bibinfo {author} {\bibfnamefont {J.~C.}\ \bibnamefont {Reyes}}, \bibinfo
  {author} {\bibfnamefont {L.~S.}\ \bibnamefont {Matthews}}, \bibinfo {author}
  {\bibfnamefont {T.~W.}\ \bibnamefont {Hyde}}, \bibinfo {author}
  {\bibfnamefont {A.}~\bibnamefont {Kovacs}},\ and\ \bibinfo {author}
  {\bibfnamefont {Z.}~\bibnamefont {Donko}},\ }\bibfield  {title} {\bibinfo
  {title} {Crystallization dynamics of a single layer complex plasma},\ }\href
  {https://doi.org/Artn 115004 Doi 10.1103/Physrevlett.105.115004} {\bibfield
  {journal} {\bibinfo  {journal} {Phys. Rev. Lett.}\ }\textbf {\bibinfo
  {volume} {105}},\ \bibinfo {pages} {115004} (\bibinfo {year}
  {2010})}\BibitemShut {NoStop}%
\end{thebibliography}%

\end{document}